# Status and performance of the ALICE Fast Interaction Trigger

**Yury Melikyan**[a,*] **on behalf of the ALICE collaboration**

[a]*Helsinki Institute of Physics*
*P.O. Box 64, Helsinki 00014, Finland*

*E-mail:* yumelikyan@gmail.com

The ALICE apparatus at CERN has undergone significant upgrade during the LHC's Long Shutdown 2 to increase the experiment's rate capability and measurement precision for Run 3 and Run 4. One of the key detectors newly introduced to the experiment is the Fast Interaction Trigger (FIT). Featuring minimal dead time, FIT provides a comprehensive trigger menu, including the minimum bias trigger with an efficiency exceeding 90% in proton-proton collisions. It also provides centrality, event plane and background measurement capabilities. Precision of the collision time measurements performed by FIT reached remarkable values of $\sigma$ = 17 ps in proton-proton collisions and $\sigma$ = 4.4 ps in Pb-Pb collisions. This paper details FIT's design, performance, and the technical novelties made to meet the stringent environmental requirements of ALICE.

---

[*]Speaker

1. Introduction

ALICE (A Large Ion Collider Experiment) has undergone a major upgrade during the Long Shutdown 2 (2019-2021) of the Large Hadron Collider (LHC) at CERN [1]. One of the key objectives for the upgrade was a significant increase in the collision rate capability of the experiment, allowing to increase the luminosity recorded by ALICE during the LHC Runs 3 and 4. The operational proton-proton collision rate in Run 3 is ~500 kHz, with the target maximum collision rate for Pb ions reaching ~50 kHz. This constitutes a factor of x3 to x6 increase compared to ALICE's rate capability during the LHC Run 2. To meet the requirements of the upgraded experiment, a new Fast Interaction Trigger (FIT) detector [2], along with several other brand new or upgraded detector subsystems [1, 3], has been designed and deployed.

2. Required functionality and design constraints

**2.1 Functionality of FIT**

In proton-proton (pp) collision mode, the LHC beams are structured in bunches of $>10^{11}$ protons each. Bunches of the two counter-directional beams intersect discretely in time with ~25 ns intervals [4]. Being the key ALICE luminometer, FIT is required to measure and transfer the actual number of bunches collided in a unit of time to the LHC side. For this, FIT must feature BC-per-BC readout capability, i.e. sub-nanosecond timing resolution and dead time below 25 ns.

The 25-ns BC interval is not precisely hold. The actual collision vertex fluctuates ±10 cm away from the interaction point along the beam axis (see Section 3). Particle vertex is precisely reconstructed by the tracking detectors, but this is done offline with a significant delay. This necessitates FIT to measure the collision time with ~30 ps time resolution, nailing down the vertex with a sub-centimeter precision in the online mode. Apart from the background monitoring, this feature allows the ALICE TOF detector to use the FIT signal as a time-zero reference [5].

Although most of the ALICE detectors are able to operate in a trigger-less mode [1], some detectors do require a trigger signal based on the collision multiplicity info. Moreover, event recording by selected subsystems requires a "wake-up" signal delivered with minimal latency [6]. That said, FIT's trigger signal shall be delivered to the input of ALICE's Central Trigger Processor (CTP) within less than 425 ns from the actual collision [2], including ~150 ns of cable delay.

Determination of centrality and event-plane orientation in heavy-ion collisions is also done based on the FIT signals. This calls for precise measurement of particles' multiplicity in the forward region, thus requiring the ability to measure particle flux in a dynamic range exceeding 1:500 per channel.

Finally, FIT provides an input to the veto signal for the selection of the Ultra-Peripheral Collisions (UPC) and participates in diffractive physics studies.

**2.2 Design constraints**

The critical design constraints limiting the choice of detector technologies were:
- Presence of strong magnetic field of flipping polarities with an induction of up to 0.45 T.
- Harsh radiation conditions. Innermost detector channels will be exposed to ~0.5 MRad dose or ~$10^{13}$ 1-MeV-$n_{eqv}$ / $cm^2$ hadron fluence equivalent for silicon detectors.

- Strict space limitations. Presence of the massive muon absorber and introduction of the brand-new Muon Forward Tracker (MFT) detector shrank the longitudinal size of the room available in the IP vicinity at the C-side to less than 90 mm.

## 3. Detector description

FIT comprises a set of three subsystems composed in five standalone detector elements:

- FIT Time-zero detector (FT0) – optimised for timing measurements;
- FIT Vertex-zero detector (FV0) – providing hermetic coverage of a large pseudorapidity range with a wide dynamic range in particle flux;
- Forward Diffractive Detector (FDD) – featuring very-forward placement for the needs of diffractive physics.

Figure 1 shows the layout of the five standalone FIT detectors. FT0 and FDD are paired detectors located on A- and C- side from the Interaction Point (IP). FV0 is located on the A-side only.

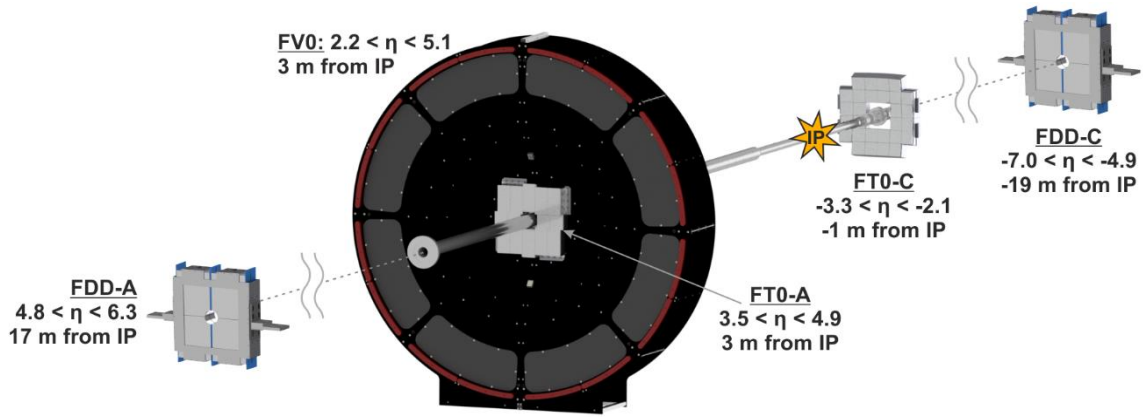

Figure 1. Layout of the FIT detectors.

### 3.1 FT0

FT0 is the only Cherenkov detector among the FIT subsystems. Timewise, it is the most precise subsystem of ALICE. The two FT0 arrays are composed of the Cherenkov modules. Each module contains four identical fused silica radiator bars with mirror-coated side walls. By means of the Dow Corning 200 optical grease, these bars are placed in a direct optical contact with a multianode microchannel plates-based photomultiplier tube (MCP-PMT). See Figure 2 for the photos of the detector modules.

As could be seen from Figure 1, the Cherenkov modules are assembled in two arrays of 24 and 28 units – FT0-A and FT0-C correspondingly – located at opposite sides of the IP. Given the asymmetric placement of the arrays, the particle load of their innermost channels differs by a factor of ~7. Table 1 shows the spread in particle load experienced by the inner and outer channels of FT0-A and FT0-C at the default pp collision rate. Table 2 shows the same parameters, but for 50 kHz Pb-Pb collisions. Note the Pb-Pb collision rate achieved by the LHC in 2022-2023 did not exceed 44 kHz, with the majority of the 2023 Pb-Pb data collected with luminosity levelling at 25 kHz equivalent rate.

Choice of the MCP-PMTs as photosensor type was driven by the combination of environmental restrictions and a chase for the best possible timing resolution. The Planacon MCP-PMT family was selected as the most suitable for the task. Significant R&D was performed to optimise them to the ALICE conditions and photon load listed in Table 2 [7, 8, 9]. The resulting MCP-PMT model is Planacon XP85002/FIT-Q, with 52 units used in FIT. It is the first massive application of Planacons in a running accelerator-based experiment.

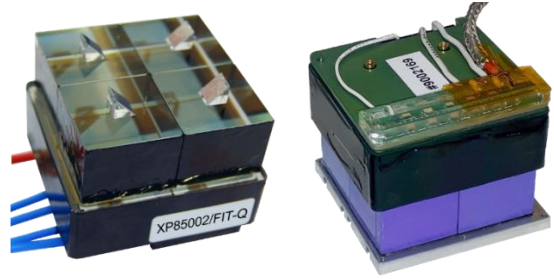

Figure 2. FT0 Cherenkov modules. The rear planes of the bars are equipped with prism-shaped optical inputs meant for detector monitoring & calibration. Residual surface was blackened after taking this photo to largely suppress the module's sensitivity to background particles flowing backwards.

Table 1. Load on the FT0 channels in pp collisions mode.

| 500 kHz pp | A-inner | A-outer | C-inner | C-outer |
|---|---|---|---|---|
| **Particle flux** (thousand MIPs/s/channel) | 1300 | 5.4 | 180 | 3.1 |
| **Photon flux** (million p.e./cm$^2$) | 54 | 2.2 | 7.4 | 0.13 |
| **Anode current** (nA/cm$^2$) | 110 | 4.5 | 15 | 0.26 |

Table 2. Load on the FT0 channels in Pb-Pb collisions mode.

| 50 kHz PbPb | A-inner | A-outer | C-inner | C-outer |
|---|---|---|---|---|
| **Particle flux** (million MIPs/s/channel) | 5.9 | 0.37 | 0.95 | 0.2 |
| **Photon flux** (million p.e./cm$^2$) | 240 | 15 | 39 | 8.3 |
| **Anode current** (nA/cm$^2$) | 490 | 31 | 78 | 17 |

### 3.2 FV0

FV0 is based on a circular array of EJ-204 plastic scintillators arranged in five rings. The rings are optimized in width and segmented to provide equal pseudorapidity coverage. Given the complex shape of the segments and the large surface of the detector plane, direct photon readout was not an option for FV0. Conventional solutions for the photon readout in such case utilise wavelength-shifting (WLS) fibres. They ensure sufficiently high light yield but worsen the time resolution achievable with the plastic scintillator. To solve the problem, FV0 was equipped with a novel light readout. It utilizes ~50 000 clear PMMA fibres (Asahi SB-1000) in direct optical contact with the rear scintillator surface (see Figure 4). Photon readout is based on 48 Hamamatsu R5924 fine-mesh PMTs, ensuring the time resolution for single minimum ionising particles (MIPs) below ~200 ps [2, 10].

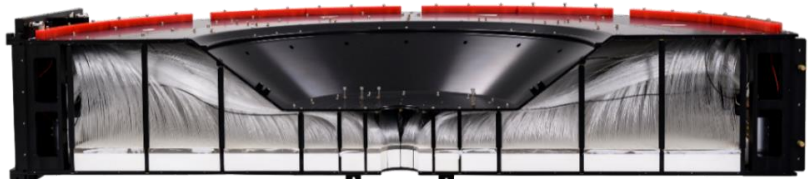

Figure 4. Cross-section view of the assembled FV0 detector.

### 3.3 FDD

Each of the two FDD detectors is composed of two layers of scintillation counters working in coincidence mode. The four counters of each layer are made of the BC-420 scintillator tiles with photon readout through NOL-38 WLS bars and clear plastic fibres to Hamamatsu R7761 fine-mesh PMTs [11].

4. Detector performance

Despite its non-hermetic design, FT0 provides efficient triggering for ALICE. It produces triggers from individual modules having signal above half of a MIP within ±2 ns from the expected colliding BCs time. FT0VX, the vertex trigger, is produced at the coincidence of FT0-A and FT0-C "or" triggers – its overall efficiency exceeds 90% for those events with the vertex reconstructed by the inner tracker (see Figure 5).

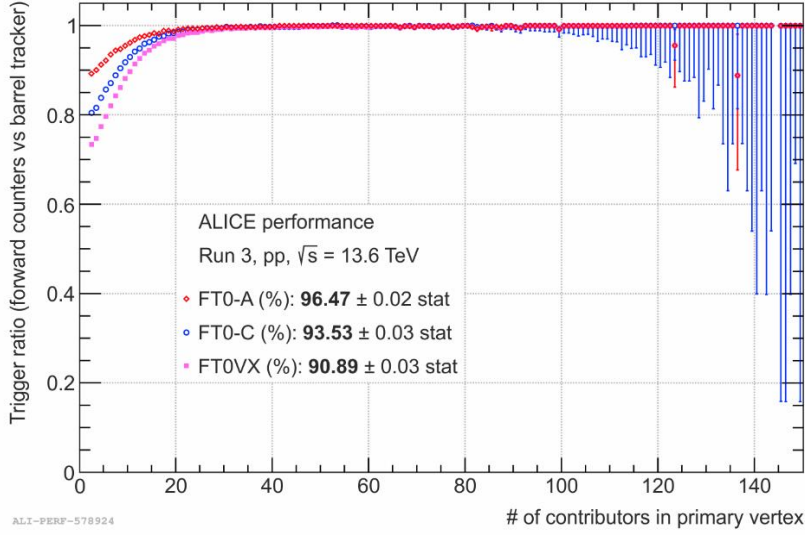

Figure 5. Fraction of events firing FT0 triggers among the events reconstructed by the Inner Tracking System of ALICE working in a trigger-less mode.

As mentioned in Section 2.1, the collision point of the LHC beams fluctuates along the beamline. The two FT0 detector assemblies measure the collision time relative to the ~40 MHz High-Quality (HQ) LHC clock [4]. The average of the signal time coming from FT0-A and FT0-C gives the collision time, and their difference gives the collision vertex. The profile of the reconstructed collision time and vertex in top-energy pp collisions is shown in Figure 6 (a), in Pb-Pb collisions – in Fig.7 (a). One can note the excellent collision time resolution FT0 is able to provide: σ = 17 ps in pp mode and σ = 4.4 ps in Pb-Pb mode (see Figure 6 (b) and 7 (b) respectively). The longitudinal vertex coordinate defined from the FT0 timing signals correlate well to that reconstructed offline with the Inner Tracking System (ITS) [1].

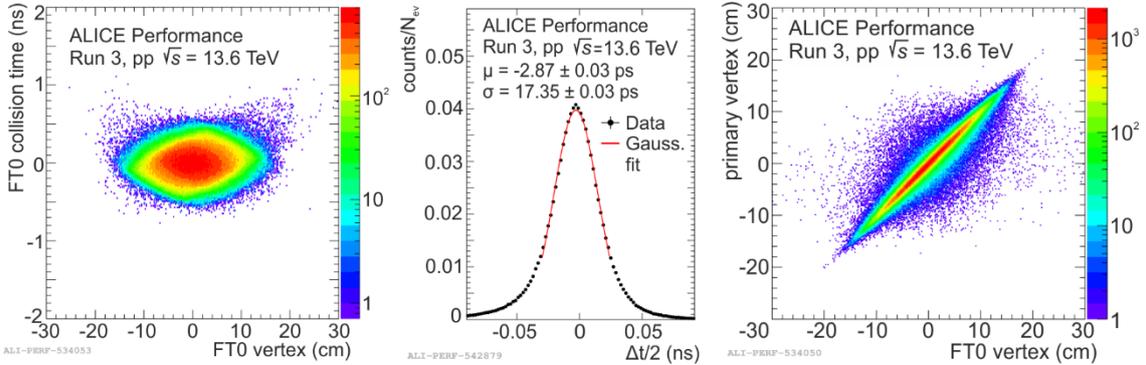

Figure 6. FIT performance in pp collisions at √s = 13.6 TeV. a) Collision time versus collision vertex reconstructed with FT0; b) FT0 collision time resolution; c) Correlation between the collision vertex reconstructed with FT0 and ITS.

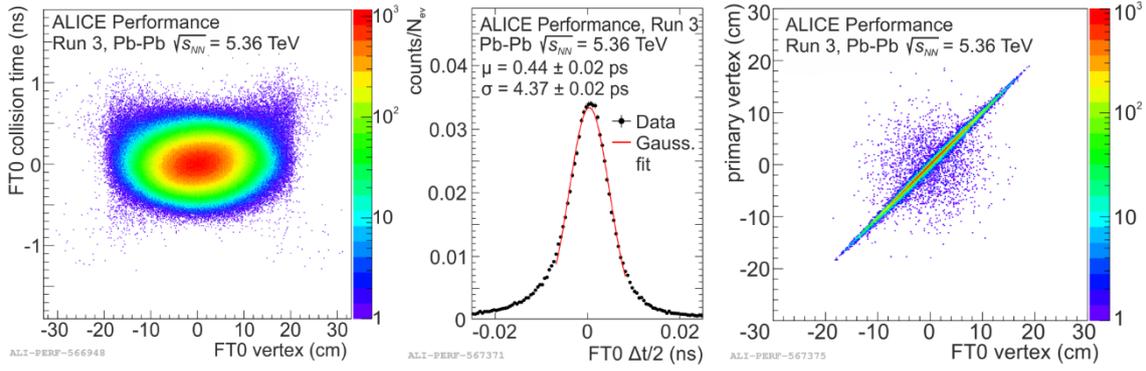

Figure 7. FIT performance in Pb-Pb collisions at $\sqrt{s_{NN}}$ = 5.36 TeV. a) Collision time versus collision vertex reconstructed with FT0; b) FT0 collision time resolution; c) Correlation between the collision vertex reconstructed with FT0 and ITS.

The rate capability of all FT0 MCP-PMTs is sufficiently high to maintain their gain-matching at 25 kHz Pb-Pb levelling within a spread of less than ±10%. Higher rates cause anode current saturation in the eight innermost FT0-A MCP-PMTs [9], while the other 44 MCP-PMTs remain gain-matched. Thus, the residual error in centrality determination does not exceed 3% for the highest Pb-Pb rates achieved at the LHC.

The integrated anode charge collected so far in the innermost FT0 photosensors exceeds 1.0 $C/cm^2$. Ageing processes in the few most loaded MCP-PMTs resulted in a factor of ~2 loss in the response function of the eight innermost FT0 photosensors, complying well with the outcome of our dedicated ageing test performed before the detector deployment [7]. No significant influence on the detector timing is expected up to a factor of ~10 loss in the MCP-PMT response function.